\documentclass[aps,prb,twocolumn,groupedaddress, showpacs]{revtex4}

\usepackage{graphicx}

\begin{document}

\title{Dilatometry study of the ferromagnetic order in
single-crystalline URhGe}

\author{S. Sakarya} \email[Corresponding author: ]{S.Sakarya@iri.tudelft.nl}
\author{N. H. van Dijk}
\affiliation{Interfaculty Reactor Institute, Delft University of
Technology, Mekelweg 15, 2629 JB Delft, The Netherlands}

\author{A. de Visser} \author{E. Br\"{u}ck}
\affiliation{Van der Waals -- Zeeman Institute, University of
Amsterdam, Valckenierstraat 65, 1018 XE Amsterdam, The
Netherlands}

\date{\today}

\begin{abstract}
Thermal expansion measurements have been carried out on
single-crystalline URhGe in the temperature range from 2 to 200 K.
At the ferromagnetic transition (Curie temperature $T_C = 9.7
$~K), the coefficients of linear thermal expansion along the three
principal orthorhombic axes all exhibit pronounced positive peaks.
This implies that the uniaxial pressure dependencies of the Curie
temperature, determined by the Ehrenfest relation, are all
positive. Consequently, the calculated hydrostatic pressure
dependence $ d T_C / d p $ is positive and amounts to 0.12 K/kbar.
In addition, the effective Gr\"{u}neisen parameter was determined.
The low-temperature electronic Gr\"{u}neisen parameter
$\Gamma_{\mbox{sf}}$ = 14 indicates an enhanced volume dependence
of the ferromagnetic spin fluctuations at low temperatures.
Moreover, the volume dependencies of the energy scales for
ferromagnetic order and ferromagnetic spin fluctuations were found
to be identical.
\end{abstract}


\maketitle

\section{Introduction}

Recently, the intermetallic compound URhGe has attracted much
attention because superconductivity ($T_c$ = 0.25 K) was found to
coexist with ferromagnetism (Curie temperature $T_C$ = 9.5 K).
\cite{1} The surprising discovery of superconductivity at ambient
pressure in this itinerant ferromagnet was preceded by the
discovery of (pressure-induced) superconductivity in the itinerant
ferromagnets UGe$_2$ \cite{2} and ZrZn$_2$. \cite{3} Until these
discoveries, it was generally believed that ferromagnetic order
excludes superconductivity. This is nicely demonstrated by
experiments on systems like ErRh$_4$B$_4$ \cite{4,5} and
HoMo$_6$S$_8$, \cite{6} where standard BCS singlet-type
superconductivity is suppressed when ferromagnetic order sets in.
The most likely explanation for the appearance of
superconductivity in these weak itinerant ferromagnets is that the
superconducting state is mediated by ferromagnetic spin
fluctuations, giving rise to Cooper pairs with parallel spins ($S
= 1$). \cite{7,8,9,10} This type of pairing is relatively
insensitive to a local magnetic field and can therefore coexist
with ferromagnetic order. The pressure-dependent experiments on
UGe$_2$ and ZrZn$_2$ suggest that in these systems
superconductivity emerges near a ferromagnetic quantum critical
point, i.e. when the ferromagnetic transition temperature is tuned
to $T_C = 0$. At the quantum critical point the ferromagnetic spin
fluctuations are strongly enhanced. One may therefore expect that
ferromagnetic order in URhGe is also very sensitive to pressure.

URhGe crystallizes in the orthorhombic TiNiSi-type structure
(space group $P_{nma}$). \cite{11} The unit cell, with dimensions
$ a = 6.87$ \AA, $b = 4.33 $ \AA, and $c = 7.51 $ \AA, contains 4
formula units. Neutron-diffraction experiments on
single-crystalline URhGe \cite{12} revealed a collinear
ferromagnetic order below $T_C$ = 9.6 K with ordered U moments of
0.35 $\mu_{\mathrm{B}}$ confined to the $b-c$ plane. No component
of the ordered moment was observed along the $a$ axis, which acts
as the hard magnetic direction for the magnetization. In addition
to the neutron-diffraction experiments, the ferromagnetic order in
single-crystalline URhGe was studied by specific-heat,
magnetization, and electrical resistivity, \cite{12,13} which
showed a sizeable influence of applied magnetic fields on the
ferromagnetic order and on the ferromagnetic spin fluctuations in
the $b-c$ plane. In the low-temperature limit, the specific heat
is characterized by the electronic contribution of the
ferromagnetic spin fluctuations with a moderately enhanced linear
term of $\gamma = 164$ mJ/molK$^2$. The magnetic properties of
single-crystalline URhGe are in good agreement with the results
from earlier measurements on polycrystalline and powder samples,
\cite{14,15,16,17,18,19} which have been reviewed by Sechovsky and
Havela. \cite{11} Recently, band-structure calculations were
performed by Divis and co-workers \cite{20} and Shick \cite{21} to
study the origin of the magnetic order in URhGe. These
calculations suggest a substantial hybridization between the
U-$5f$ and Rh-$4d$ states and a relatively small Uranium magnetic
moment of 0.3 $\mu_{\mathrm{B}}$ due to a partial cancellation of
the spin and orbital components. The calculated moments are in
good agreement with the measured values.

In this paper we report thermal expansion measurements of
single-crystalline URhGe in the temperature range from $T$ = 2 to
200 K. Our principal aim was to determine the pressure dependence
of the ferromagnetic transition temperature $T_C$. For a
second-order phase transition, the uniaxial pressure dependence of
$T_C$ (at ambient pressure) can be determined with the Ehrenfest
relation from the anomalies in the linear coefficient of thermal
expansion and the specific heat. The initial pressure dependence
may give an estimate of the critical pressure needed to suppress
the ferromagnetic order and reach the quantum critical point at
$T_C$ = 0. In addition, we have determined the electronic
Gr\"{u}neisen parameter, which characterizes the volume dependence
of the ferromagnetic spin fluctuations at low temperatures.

\section{Experimental}

The dilatometry experiments were performed on a single-crystalline
URhGe sample with dimensions $a \times b \times c = 2.4 \times 5.0
\times 2.4$ mm$^3$. The sample was cut from the material used in
earlier specific-heat measurements performed by Hagmusa and
co-workers. \cite{13} The crystal has been grown from a
stoichiometric melt of at least 99.95 \% pure materials by means
of a modified tetra-arc Czochralski technique in a continuous
gettered Ar atmosphere. No subsequent heat treatment was given to
the crystal. Due to the relatively high residual resistivity at
low temperatures no superconductivity was observed in this
particular crystal. \cite{12} The coefficient of linear thermal
expansion $\alpha (T) = (1/L)(dL/dT)$ was measured using a
sensitive parallel-plate capacitance dilatometer \cite{22} along
the orthorhombic $a$, $b$, and $c$ axis of the crystal. From these
measurements the volume expansion $\alpha_v = \alpha_a + \alpha_b
+ \alpha_c$ was determined.

\section{Results}

\begin{figure} \resizebox{8cm}{!}{
\includegraphics{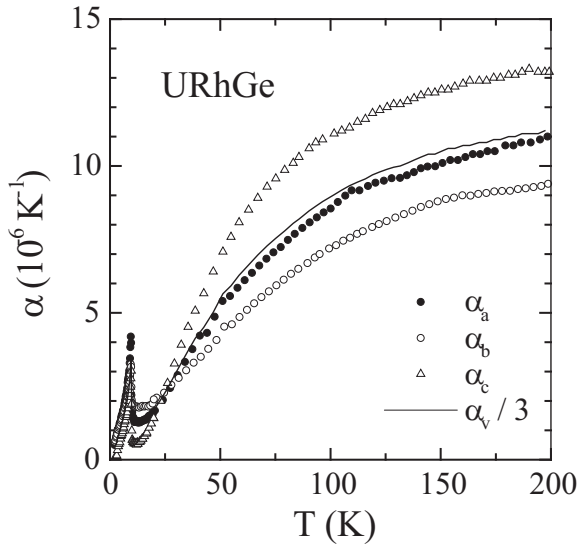} } \caption{\label{figT=2to200K}The
coefficients of linear thermal expansion $\alpha$ of URhGe as a
function of temperature $T$ along the orthorhombic $a$, $b$, and
$c$ axis. For comparison the volume expansion divided by a factor
three ($\alpha_v/3$) is also shown. The high-temperature behavior
is governed by the phonon contribution, while the anomaly at $T_C$
= 9.7 K reflects the onset of ferromagnetic order. } \end{figure}

In Fig. \ref{figT=2to200K} 
the coefficient of linear thermal expansion $\alpha$ along the
$a$, $b$, and $c$ axis of single-crystalline URhGe is shown as a
function of temperature in the range from $T$~=~2 to 200 K. The
temperature dependence of the volume expansion $\alpha_v =
\alpha_a + \alpha_b + \alpha_c$ is shown for comparison (notice
the figure shows $\alpha_v/3$). At high temperatures the thermal
expansion is governed by the phonon contribution for all three
orientations. Around a temperature of 25 K a remarkable crossing
of the curves for the thermal expansion along the $a$, $b$, and
$c$ axis is observed. This crossing is a clear sign for the
development of an additional contribution from ferromagnetic spin
fluctuations at low temperatures. It is interesting to note that
this additional contribution from ferromagnetic spin fluctuations
mainly affects the anisotropy of the thermal expansion in the
$b-c$ plane, which acts as the easy plane for the magnetization.
At $T_C$ = 9.7 K the ferromagnetic order sets in and a peak in the
coefficient of linear thermal expansion is observed for all three
directions.

\begin{figure} \resizebox{8cm}{!}{
\includegraphics{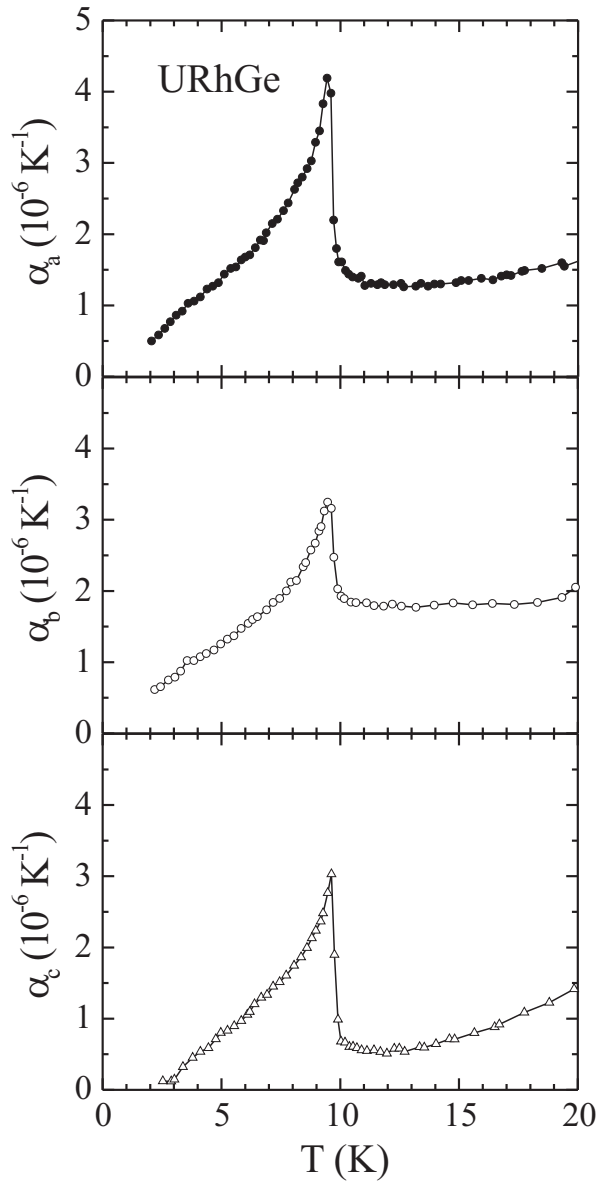} } \caption{\label{figT=2to20K}The
coefficients of linear thermal expansion $\alpha$ of URhGe as a
function of temperature $T$ along the orthorhombic $a$, $b$, and
$c$ axis at low temperatures. The anomaly at $T_C$~=~9.7 K
reflects the onset of ferromagnetic order. The large linear term
at low temperatures is due to spin fluctuations. }
\end{figure}

The low-temperature behavior of the coefficients of linear thermal
expansion along the $a$, $b$, and $c$ axis is shown in more detail
in Fig. \ref{figT=2to20K}.  The observed steps in the coefficients of linear
thermal expansion have the same sign but different sizes for the
three orthorhombic axes of single-crystalline URhGe. The values of
the steps are listed in Table \ref{steps}. \begin{table}
\caption{\label{steps}Step anomalies in the coefficients of linear
thermal expansion of single-crystalline URhGe along the
orthorhombic $a$, $b$, and $c$ axis. The corresponding pressure
dependence of the Curie temperature, $d T_C / d p$, was deduced
from the Ehrenfest relation (see text). }
\begin{ruledtabular} \begin{tabular}{ccc} & $ \Delta\alpha $ &
$dT_C/d p$ \\ & $ 10^{-6} $ K$^{-1}$ & K/kbar \\
\hline $a$ axis & 3.4(1) & 0.052(3) \\ $b$ axis & 1.7(1) &
0.026(2) \\ $c$ axis & 2.7(1) & 0.041(2) \\ volume   & 7.8(2) &
0.119(6) \end{tabular} \end{ruledtabular} \end{table} In Fig.
\ref{figmisc0_30K} the low-temperature volume expansion divided by
temperature $\alpha_v / T$ is shown as a function of temperature
and compared with the specific heat divided by temperature, $c/T$,
measured on a sample prepared from the same single-crystalline
batch. \cite{13}

\section{Discussion}

\begin{figure} \resizebox{8.2cm}{!}{
\includegraphics{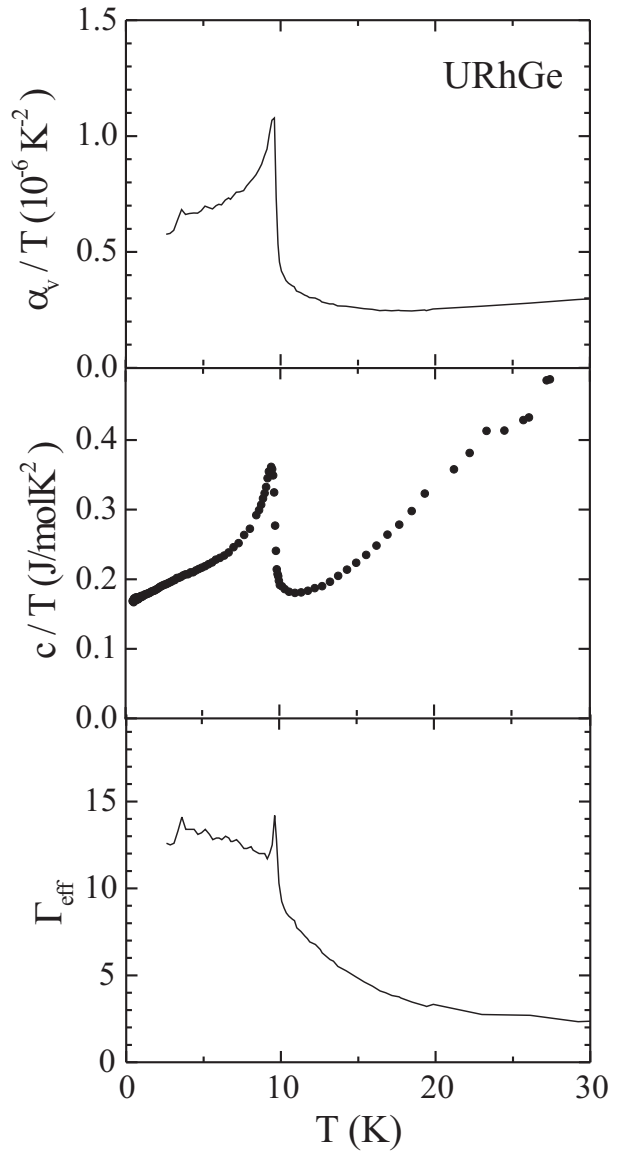} } \caption{\label{figmisc0_30K}The
volume expansion divided by temperature $\alpha_v /T $ of URhGe as
a function of temperature $T$ at low temperatures. For comparison
the specific heat divided by temperature, $c/T$, of a sample
prepared from the same single-crystalline batch is shown.
\cite{13} The bottom frame shows the effective Gr\"{u}neisen
parameter $\Gamma_{\mathrm{eff}}$, determined from the
experimental data of the volume expansion and the specific heat
(see text). } \end{figure}

The temperature dependence of the thermal expansion at high
temperatures is governed by the phonon contribution and closely
resembles a Debye curve. The estimated Debye temperature of
$\theta_D \approx 200$ K is in good agreement with the specific
heat data. \cite{12,13} At low temperatures the Debye curve for
the phonon contribution is expected to show a $T^3$ temperature
dependence. Below $T = 30$ K the thermal expansion along the $a$
axis shows a clear deviation of this behavior, which is even more
pronounced along the $b$ axis. This deviation indicates the
development of an additional contribution due to ferromagnetic
spin fluctuations. This additional contribution was also observed
in the specific heat measurements and described in terms of an
enhanced electronic contribution. \cite{12} In the
ferromagnetically ordered state below $T_C = 9.7$~K the
temperature dependence of the volume expansion closely resembles
that of the specific heat, as shown in Fig. \ref{figmisc0_30K}. In
line with the analysis of the specific heat measurements \cite{12}
three different contributions to the thermal expansion can be
identified in the ferromagnetically ordered state, namely
contributions due to phonons, ferromagnetic spin waves, and
ferromagnetic spin fluctuations. As discussed, the phonon
contribution shows a $T^3$ power-law behavior at low temperatures.
The ferromagnetic spin-wave contribution is expected to obey a
$T^{3/2}$ power-law behavior, while the ferromagnetic spin
fluctuations lead to an enhanced linear term at low temperatures.
Both the phonon and spin-wave contributions to the volume
expansion divided by temperature $\alpha_v/T$ (and the specific
heat divided by temperature $c/T$) vanish at low temperatures and,
as a consequence, the extrapolated value of $\alpha_v / T = 5.8(2)
\times 10^{-7} K^{-2}$ at $T = 0$ is solely due to the
contribution of the ferromagnetic spin fluctuations. As expected
for an itinerant ferromagnetic system, no indications of a crystal
field contribution were observed in the temperature dependence of
the coefficients of linear thermal expansion.

In order to determine the uniaxial and hydrostatic pressure
dependence of the ferromagnetic transition temperature we have
applied the Ehrenfest relation. For a second order phase
transition, the uniaxial pressure dependence of the transition
temperature is directly related to the step anomalies in the
coefficient of linear thermal expansion and the specific heat by
the Ehrenfest relation: \begin{equation} \label{eq1} \frac{d
T_C}{d p_i} = \frac{ V_m \Delta\alpha_i}{\Delta \left( c / T
\right)}, \end{equation} where the index $i$ refers to the
orthorhombic axis, $V_m = 3.36 \times 10^{-5}$ m$^3$/mol is the
molar volume and $\Delta \left( c / T \right)$ = 0.22(1)
J/molK$^2$ is the anomaly in the specific heat divided by
temperature. \cite{12} By applying this relation to the
experimental step anomalies in the coefficients of linear thermal
expansion, the uniaxial pressure dependence of $T_C$ along the
$a$, $b$, and $c$ axis is obtained. The calculated values are
listed in Table \ref{steps}. The hydrostatic pressure dependence
of $T_C$ can be obtained by inserting the volume expansion for the
coefficient of linear expansion in Eq. (\ref{eq1}), or by summing
the three contributions of the uniaxial pressure dependence. The
different pressure dependencies of $T_C$ as listed in Table
\ref{steps} are all positive. This strongly suggests that the
ferromagnetic order cannot be suppressed by moderate mechanical
hydrostatic or uniaxial pressures, like in the case of UGe$_2$ and
ZrZn$_2$. Instead a \textit{negative} uniaxial pressure is needed
to suppress $T_C$ for all crystallographic directions. Using a
simple linear extrapolation of the initial pressure dependence
calculated from the Ehrenfest relation, we arrive at a negative
critical hydrostatic pressure of $p_{cr} \approx -80$ kbar. It is
important to note that this value should be regarded as an upper
bound for the negative critical pressure as the pressure
dependence of $T_C$ is expected to show significant non-linear
corrections to the initial pressure dependence at ambient
pressure. A negative critical pressure of the order of $p_{cr}
\approx -80$ kbar might be achieved by suitable chemical
substitutions.

In order to characterize the volume dependence of the electron
correlations we have calculated the effective Gr\"{u}n\-eisen
parameter of URhGe. The effective Gr\"{u}n\-eisen parameter
$\Gamma_{\mathrm{eff}}$ is calculated from the
temperature-dependent volume expansion $\alpha_v(T)$ and specific
heat~$c(T)$:
\begin{equation} \Gamma_{\mathrm{eff}} (T) = \frac{ V_m \alpha_v (T) }{
\kappa c(T) } \end{equation} where $\kappa$ = -(1/$V$)($dV/dp$) is
the isothermal compressibility. As the compressibility of URhGe is
unknown we will use an estimated value $\kappa = 0.8$ Mbar$^{-1}$.
Experimental values for the compressibility of other UTX compounds
vary from $\kappa = 0.6$ to $1.0$ Mbar$^{-1}$. \cite{23}

In Fig. \ref{figmisc0_30K}  the effective Gr\"{u}neisen parameter
$\Gamma_{\mathrm{eff}}$, calculated from the experimental volume
expansion and the reported specific heat \cite{13}, is shown as a
function of temperature. At high temperatures the effective
Gr\"{u}neisen parameter shows a small constant value of
$\Gamma_{\mathrm{ph}} = 2$ and describes the volume dependence of
the characteristic energy scale for the phonons. Below 30~K the
effective Gr\"{u}neisen parameter rapidly increases and reaches a
value of $\Gamma_{\mathrm{eff}} \approx 14$ just above $T_C$.
Below $T_C$ a weak suppression of the effective Gr\"{u}n\-eisen
parameter is observed with a slow increase for decreasing
temperatures. In the low temperature limit $\Gamma_{\mathrm{eff}}$
corresponds to the enhanced electronic Gr\"{u}neisen parameter
$\Gamma_{\mathrm{sf}} = d \ln \gamma / d \ln V \approx 14$ of the
ferromagnetic spin fluctuations. The corresponding relative
pressure dependence of the electronic specific heat amounts to $d
\ln \gamma / d p = -\kappa \Gamma_{\mbox{sf}} \approx -18 $
Mbar$^{-1}$.

The relation between magnetic order and the spin fluctuations can
further be studied by comparing the volume dependence of the
energy scales for the ferromagnetic order ($T_C$) and the
ferromagnetic spin fluctuations ($T_{\mathrm{sf}}$). It turns out
that the Gr\"{u}neisen parameter for the ferromagnetic order
$\Gamma_{\mathrm{F}} = - d \ln T_C / d \ln V = (1 / \kappa T_C) (d
T_C / d p) \approx 16$ is of the same order of magnitude and has
the same sign as the Gr\"{u}neisen parameter for the ferromagnetic
spin fluctuations $\Gamma_{\mathrm{sf}} = - d \ln T_{\mathrm{sf}}
/ d \ln V = d \ln \gamma / d \ln V \approx 14$. This situation is
in strong contrast to pressure-induced antiferromagnetic
superconductors like CePd$_2$Si$_2$ \cite{24} where the
antiferromagnetic order competes with the spin fluctuations with
an opposite scaling behavior with volume. It can therefore be
expected that the spin-mediated superconductivity of URhGe exists
over a wide pressure range, as observed for the ferromagnetic
superconductor ZrZn$_2$. \cite{3} This in contrast to the
situation in the ferromagnetic superconductor UGe$_2$, \cite{2}
where superconductivity is only observed in a small pressure
region close to the critical pressure where the ferromagnetic
order is suppressed.

\section{Conclusions}

We have performed thermal expansion measurements on a
single-crystalline sample of the ferromagnet URhGe. Below the
ferromagnetic ordering temperature of $T_C = 9.7$ K an increase in
the coefficient of linear thermal expansion was observed along all
three orthorhombic axes. The uniaxial pressure dependence of the
ferromagnetic transition temperature was determined by the
Ehrenfest relation from the anomalies in the coefficients of
linear thermal expansion and the specific heat. We find positive
values of $d T_C / d p$ for all principal axes. Consequently, the
hydrostatic pressure dependence is also positive and amounts to $d
T_C / d p$ = 0.12 K/kbar. This contrasts the behavior reported for
UGe$_2$ and ZrZn$_2$. In addition, the effective Gr\"{u}neisen
parameter was determined. The resulting low-temperature behavior
points to an enhanced volume dependence of the ferromagnetic spin
fluctuations at low temperatures and an equal volume scaling of
the energy scales for the ferromagnetic order and the
ferromagnetic spin fluctuations.

\begin{acknowledgments}
We thank B. F\aa k and P.C.M. Gubbens for stimulating discussions.
\end{acknowledgments}

\end{document}